# µeV-deep neutron bound states in nanocrystals

Hao Tang [1], Guoqing Wang [2,3], Paola Cappellaro [2,3], Ju Li [3,1,*]

[1] Department of Materials Science and Engineering, Massachusetts Institute of Technology, Cambridge, MA 02139, USA

[2] Research Laboratory of Electronics, Massachusetts Institute of Technology, Cambridge, MA 02139, USA

[3] Department of Nuclear Science and Engineering, MIT, Cambridge, MA 02139, USA

* To whom correspondence should be addressed: liju@mit.edu

***Abstract***: *The nuclear strong force induces the widely studied neutron scattering states and MeV-energy nuclear bound states. Whether this same interaction could lead to low-energy bound states for a neutron in the nuclear force field of a cluster of nuclei is an open question. Here, we computationally demonstrate the existence of -µeV-level neutronic bound states originating from nuclear interaction in nanocrystals with a spatial extent of tens of nanometers. These negative-energy neutron wavefunctions depend on the size, dimension, and nuclear spin polarization of the nanoparticles, providing engineering degrees of freedom for the artificial neutronic "molecule".*

## Introduction

Neutron scattering is a widely studied technique to characterize materials' structure and dynamics [1]. In neutron scattering, interference of neutron's positive-energy scattering states in a cluster of atoms is utilized to probe the atomic configuration [2], magnetic structures [3], and ionic motion [4]. Besides scattering states, the strong nuclear interaction between neutron and nuclei can also trap a neutron in femtometer-scale bound states, known as the radiative neutron capture [5]. Such bound states have MeV-scale binding energy, where the neutron and nucleus combine into a new isotope and emit one $\gamma$-ray photon [6]. The energy spectra of the neutron, therefore, include the continuum spectra from scattering states covering the positive energy range and the discrete lines from bound states with ~MeV deep negative energy. The energy gap between the deep negative and positive energy spectra contains no bound states [7] if the neutron just interacts with a single nucleus, due to the short-range nature of the strong nuclear interaction [8,9]. However, it remains unknown whether a neutron interacting with a cluster of nuclei can have low-energy bound states. Intuitively, the neutron eigenstates with each nucleus interact with each other and form new eigenstates, in analogy to how the linear combination of atomic orbitals (LCAO) forms molecular orbitals [10]. It is, therefore, intriguing to probe whether there are long-lived discrete weakly bound states of neutrons, localized around a collection of atoms and ions, e.g. nanoparticles and nanowires. We define the negative-energy neutronic states with a 10nm-scale broadening the "molecular neutronic" states. Such weakly bound neutron states, if they exist, would provide a platform for designing neutron eigenfunction by controlling atomic configuration and for probing the strong nuclear interaction by low-energy neutrons. For example, the molecular neutronic states open new possibilities in probing nuclei's neutron scattering and absorption cross sections [11], neutron electric dipole moment [12,13], as well as neutron bound-state $\beta^-$-decay [14] under the low-energy limit, which contains critical information about the nuclear force.

In this work, we demonstrate the existence of the molecular neutronic state in hydrogen-containing nanocrystals by analytical derivation and computational simulations. Essentially, the multi-center superposition of positive-energy scattering states can form a *negative*-energy weakly bound state. Different from the MeV neutron bound states whose properties are set by the fixed isotope properties [6], the energy levels and wave functions of molecular neutronic states can be engineered by the host nanocrystal's size and shape. As the molecular neutronic state is similar to the electronic state in quantum dots [15], we call the system hosting such states "neutronic quantum dot" (NQD).

### *Theory*

We use direct numerical calculations and the Green function formalism to demonstrate the existence of low-energy bound states and calculate the binding energy and eigenfunction of molecular neutronic states. The neutron moves in a nuclear force potential $V(\vec{r}) = \sum_i v_i(\vec{r})$, where $v_i(\vec{r})$ is the potential of the $i$th nucleus located at $\vec{R}_i$ ($i = 1, 2, \cdots, N$ with $N$ nuclei in the system). Solving for neutron states in the NQD encounters a multiscale challenge [16]: the nuclear force $v_i$ is localized to femtometer length scale, while the interatomic distance $\vec{R}_i - \vec{R}_j$ is in the length scale of Å, exhibiting a separation of 5 orders of magnitude. That makes it prohibitively difficult to directly discretize the quantum mechanics equation on a spatial grid. In order to bypass this problem, we use the Green function formalism to show that the concept of Fermi pseudo potential [17] used in neutron scattering can also be applied to the molecular neutronic states, which encodes the fm-scale features into the scattering length [11].

The bound-state wavefunction $\psi(r)$ with an eigenenergy $E < 0$ can be obtained from an integral equation [18]

$$\psi(\vec{r}) = -\int \frac{2m_\mathrm{n}}{\hbar^2} V(\vec{r}') G(\vec{r}, \vec{r}'; E) \psi(\vec{r}') dr'^3 \tag{1}$$

over the single neutron Green function $G(\vec{r}, \vec{r}'; E) = \frac{e^{-\kappa|\vec{r}-\vec{r}'|}}{4\pi|\vec{r}-\vec{r}'|}$ satisfying $[\kappa^2 - \nabla^2]G(\vec{r}, \vec{r}'; E) = \delta(\vec{r} - \vec{r}')$, with $\kappa = \sqrt{-2m_\mathrm{n}E/\hbar^2}$, $m_\mathrm{n}$ the neutron mass and $\hbar$ the reduced Planck constant. As the support of $V(\vec{r})$ is localized to the nuclei positions, the integral can be rewritten as a sum of local integrals around each nucleus. Depending on whether $\vec{r}$ is close to a nucleus, the integral equation (1) can be rewritten as

$$\tag{2}$$

$$\psi(\vec{r}) \simeq \begin{cases} -\sum_i G(\vec{r}, \vec{R}_i; E) \int_{\Omega_i} \frac{2m_\mathrm{n} v_i(\vec{r}')}{\hbar^2} \psi(\vec{r}') dr'^3, & (\forall i \; |\vec{r} - \vec{R}_i| \gg \text{fm}) \\ -\sum_{i \neq j} G(\vec{r}, \vec{R}_i; E) \int_{\Omega_i} \frac{2m_\mathrm{n} v_i(\vec{r}')}{\hbar^2} \psi(\vec{r}') dr'^3 - \int_{\Omega_j} G(\vec{r}, \vec{r}'; E) \frac{2m_\mathrm{n} v_j(\vec{r}')}{\hbar^2} \psi(\vec{r}') dr'^3, & (|\vec{r} - \vec{R}_j| \sim \text{fm}) \end{cases}$$

Here $\Omega_i$ is a spherical volume around the $i$th nucleus with a radius of the nuclear force range. In the first case, the position $\vec{r}$ is far from all nuclei; in the second case, the position $\vec{r}$ is ~fm close to $R_j$. For all nuclei $i$ subject to $|\vec{r} - \vec{R}_i| \gg$ fm, we applied the Green function approximation $G(\vec{r}, \vec{r}'; E)|_{\vec{r}' \in \Omega_i} \simeq G(\vec{r}, \vec{R}_i; E)$. The detailed behavior of the wavefunction around the nuclei (including fm-scale oscillations) is uninfluential to the Å-scale spatial distribution of low-energy neutron states [17]. Therefore, we coarse-grain the wavefunction over an intermediate length scale fm $<< D <<$ Å and we introduce the average wavefunction, $\bar{\psi}(\vec{r}) \equiv \frac{3}{4\pi D^3} \int_{|\vec{r} - \vec{r}'| < D} \psi(\vec{r}') dr'^3$. The nuclei's influence on the average wave function can be described by the Fermi pseudo potential, $v_i^{\mathrm{PP}}(\vec{r}) = \frac{2\pi\hbar^2}{m_\mathrm{n}} \mathrm{Re}[b_i] \delta^D(\vec{r} - \vec{R}_i)$ (with $\delta^D(\vec{r}) = \frac{3}{4\pi D^3}$ if $|\vec{r}| < D$ and 0 elsewhere) [17], whose strength is characterized by the real part of the scattering length $b_i$ [11]. We prove (see supplementary information (SI) section I for details) that the following equation of the average wavefunction can be derived by integrating Eq. (2):

$$\bar{\psi}(\vec{r}) \simeq -\sum_{i, |\vec{r} - \vec{R}_i| \gg \text{fm}} \frac{e^{-\kappa|\vec{r} - \vec{R}_i|}}{|\vec{r} - \vec{R}_i|} \mathrm{Re}[b_i] \bar{\psi}(\vec{R}_i), \tag{3}$$

where the second term in the second case of Eq. (2) is proved negligible after integration. The scattering length is more frequently used to describe low-energy (compared to MeV) neutron scattering, where the neutron state has a near-zero, positive energy. In our neutron bound state case, the neutron has a near-zero negative energy. The two situations share the same scattering length (see SI section I for details). The average wavefunction at nucleus positions $\vec{R}_i$, $\bar{\psi}_i \equiv \bar{\psi}(\vec{R}_i)$ can then be obtained by solving an eigenvalue problem

$$\bar{\psi}_i + \sum_{j \neq i} \frac{e^{-\kappa|R_i - R_j|}}{|R_i - R_j|} \mathrm{Re}[b_j] \bar{\psi}_j = 0, \tag{4}$$

simultaneously determining the wavevector $\kappa$ and the binding energy $E_b \equiv \frac{\hbar^2 \kappa^2}{2m_\mathrm{n}}$. The molecular neutronic state exists if and only if Eq. (4) has non-zero solution with $\kappa > 0$. Provided that a nucleus has a negative scattering length [11], the condition can be satisfied when the size and density of the nuclear cluster exceed a threshold, thus indicating the existence of a bound state. Because $\left| \frac{e^{-\kappa R_{ij}}}{R_{ij}} \mathrm{Re}[b_j] \right|$ is far smaller than 1, the equation can be satisfied only when the summation is over a large number of nuclei, so that the second term can cancel the first term. Detailed derivation and numerical analysis are elaborated in the SI section I.

***Energy Level and Eigenfunction***

The existence of neutronic bound states thus requires negative scattering lengths, representing attractive forces to neutrons. Protons have a negative neutron scattering length with the largest magnitude among all isotopes when their nuclear spin is polarized opposite to the neutron [11]. Polarization of the nuclear spin of hydrogen nuclei can be achieved by various experimental techniques, including dynamic nuclear polarization (DNP) [19,20] and optical pumping [21]. In the following simulation, we assume all hydrogen nuclear spins in the nuclear cluster are polarized in the same direction. We use LiH nanocrystal (Fig. 1a), a widely studied hydrogen storage material [22-24], as an exemplary system to demonstrate the existence of weakly bound neutronic states by solving Eq. (4) numerically. The nanocrystalline quantum dot (nanocrystal) is assumed to have a spherical shape with a radius $R$ of tens of nanometers [25,26]. Both Li and H have an attractive nuclear force with neutrons [26], creating the negative nuclear strong-force potential shown in Fig. 1b. The binding energy levels of bound states are then calculated as a function of the nanocrystal radius, as shown in Fig. 1c. The existence of bound states requires the nanocrystal radius $R$ to be larger than a critical value, 13 nm in the LiH case. Intuitively, that is because confining neutrons in a smaller-$R$ NQD requires higher wavenumbers and thus kinetic energy, which makes the overall energy positive, so the bound states can no longer exist. Larger $R$ gives rise to multiple bound states with different symmetries (see Fig. 1d), whose binding energies all increase monotonically with $R$. The neutronic $d$ orbitals of LiH NQD split into $t_{2g}$ (the 1d (*3, meaning 3-fold degeneracy) curve) and $e_g^*$ orbitals (the 1d (*2) curve) because the cubic lattice breaks the SO(3) symmetry of the spherically shaped nanocrystal [27]. The eigenfunctions corresponding to the first two energy levels, 1s and 1p, are plotted in Fig. 1d. These neutronic eigenfunctions cover the whole nanocrystal and extend tens of nanometers into the vacuum. Moreover, the bound states and their transition frequencies can be engineered by the size and shape of the nanocrystal, providing additional tunability in quantum applications.

The binding energies of the molecular neutronic states depend on the size and dimensionality of the nanocrystal. The $\Gamma$-point neutron bound-state energy levels in the zero-dimensional LiH nanoparticle, one-dimensional LiH nanowire, and two-dimensional LiH thin film are shown in Fig. 2(a,b,c), respectively. Multiple µeV-level bound states exist and exhibit stronger binding for larger system sizes (diameter for nanoparticle and nanowire, thickness for thin film) in all three systems. Systems with higher dimensions have a smaller minimal size to host a bound state and approach stronger binding at the same system size. Different from the nanoparticle, the thin film with arbitrarily small thickness hosts bound states. That means the neutron bound state can exist in two-dimensional systems with atomic-scale thickness. The neutron bound states around the $\Gamma$ point in three-dimensional LiH perfect crystal have a parabolic band dispersion, as shown in Fig. 2d. The band structure is calculated by both Eq. (4) and plane wave basis expansion, showing consistent results. The Γ-point energy level is -0.33 μeV, which is the lower bound of the neutron energy levels in LiH. In the nanowire and thin film (Fig. 2(e,f)), the bound state energy band splits into a series of sub-bands due to the quantum confinement effect, showing the same behavior with electron band structures in low-dimensional structures.

Being spin-1/2 Fermions like electrons, the many-neutron wavefunction $\Psi(\mathbf{x}_n^1, \mathbf{x}_n^2, ...., \mathbf{x}_n^N)$ of $N$ identical neutrons, with $\mathbf{x}_n^i$ labeling both the position and spin of a neutron, must satisfy $\Psi(.., \mathbf{x}_n^i, ..., \mathbf{x}_{n'}^j, ..) = -\Psi(.., \mathbf{x}_n^j, ..., \mathbf{x}_{n'}^i, ..)$. The independent particle picture, an approximation of $\Psi(\mathbf{x}_n^1, \mathbf{x}_n^2, ...., \mathbf{x}_n^N)$, suggests that neutrons can fill up the NQD states, in a "neutronic shell model" akin to the electronic shell in molecules, with the single-particle energy and degeneracy illustrated in Fig. 1c and Fig. 2. The ground-state wavefunction is thus approximated by a Slater determinant of the $N$ lowest-energy NQD spin-orbital states. With the QD size increasing to infinity, turning the nanostructure into a bulk material, we have computed that if all the bound states are occupied by neutrons all the way to $E$=0⁻, there will be a maximal mass gain of the LiH by $6.8 \times 10^{-6}$ % (68 ppb) that should be measurable experimentally. Also, unlike electrons, the neutron-neutron interaction between these delocalized NQD spin-orbitals is rather weak, thus the many-neutron quantum state may be a good approximation of the non-interacting limit of a many-Fermion system and have some unique characteristics as a quantum information platform.

### *Neutron Absorption Lifetime*

The neutron bound states trapped in materials are intrinsically unstable because of materials' neutron absorption [28]. Besides binding energy, the lifetime is also an important feature to characterize the neutron bound state. Although neutrons interact weakly with the environment, their lifetimes are limited by the neutron absorption of H and Li nuclei, that is, the probability that the weakly bound neutron is finally absorbed by H or Li to form D or $^7$Li. For a given material (infinite crystal), we define the ground-state binding energy as $E_b^*$ and the ground-state neutron absorption lifetime as $T^*$. They can be calculated as materials properties:

$$E_b^* = -\frac{2\pi\hbar^2}{m_n\Omega} \sum_i n_i \text{Re}[b_i]$$

$$\frac{1}{T^*} = \lim_{E_k \to 0} \frac{1}{\Omega} \sum_i n_i \sigma_a^i(E_k) \sqrt{\frac{2E_k}{m_n}} \qquad (5)$$

where $\Omega$ is the unit cell volume, $n_i$, $b_i$, and $\sigma_a^i(E_k)$ are the number of $i$-type atoms in a unit cell, their scattering length, and neutron absorption cross section for neutron with a kinetic energy of $E_k$. Finite-size nanocrystal gives smaller binding energy $E_b$ and longer neutron absorption lifetime $T$, but their product is bounded by an inequality (see SI part II for derivations):

$$E_b T \leq E_b^* T^* = -\frac{\hbar}{2} \frac{\text{Im}[\sum_i n_i b_i]}{\text{Re}[\sum_i n_i b_i]} \quad (6)$$

showing that increasing the neutron absorption lifetime inevitably leads to a decrease in binding energy within the same material.

The $E_b^*$ and $T^*$ of different crystals are shown in Fig. 2, where the gray dots list all non-radioactive stable crystal structures from the Materials Project database that contain hydrogen and can host bound states [29]. The binding energies are at the level of $\mu eV$, corresponding to a required milli-kelvin-level temperature, that is already realizable in the present ultracold neutron technology [30]. The lifetime is at the level of 0.1~1 ms. In general, there is a trade-off between binding energy and lifetime. Materials satisfying Pareto optimality [31] (that means no material simultaneously has larger binding energy and longer lifetime than each selected material) are labeled by the blue points in Fig. 2, forming a frontier curve of possible $(E_b^*, T^*)$ combinations. A series of common hydrogen storage nanomaterials [32], including $MgH_2$ (0.27 $\mu eV$, 0.19 ms) and $LiBH_4$ (0.27 $\mu eV$, 0.19 ms) also exhibit reasonably high $E_b^*$ and $T^*$.

### Quantum Control

Besides the ground-state properties of the molecular neutronic states, it is also interesting to explore methods to control the transition between the ground and excited states. The neutronic states have a weak coupling to external electromagnetic fields due to the charge neutrality of neutrons. Although such a weak coupling makes it difficult to apply direct microwave control protocols to transition between different neutronic states, we propose methods to manipulate the states through indirect coupling. As the neutron bound states are sensitive to the nuclide positions, which in turn are sensitive to electromagnetic waves if the QD is charged, they can be indirectly controlled by microwave driving through nuclear force interactions. We take the LiH nanocrystal 1s and 1p neutronic states in Fig. 1d as an example to illustrate allowed dipole transition, as they can be used as the two states of a qubit controlled by microwave. The direct Zeeman interaction of a neutron spin with the microwave's magnetic field is as weak as 10 kHz under a typical experimental condition with $B \sim 10$ Gauss (for example, a 10 Gauss magnetic field shown in Ref. [33]). The corresponding Rabi oscillation time period is as long as the lifetime of neutron bound states, making microwave control through the magnetic field difficult. To achieve a strong driving, we instead propose to use an electric field, as shown in Fig. 3a. The nanocrystal is electrostatically charged by the standard charging methods [34,35], i.e. tuning the redox voltage so the net number of electrons does not balance the net nuclear charge, to a voltage of the order of magnitude of 1 V. The electric field of the microwave would then drive an oscillatory translation of the nanocrystal with net monopolar charge $q \neq 0$ and mass $M$, as in a driven oscillator model (off-resonance) [36]. Since the neutronic state is aware of the translation of the center-of-mass of the nanoparticle, this controls the time-dependent Hamiltonian for the neutron parametrically and thus can drive the Rabi oscillation of a neutron between two bound states.

The Rabi frequency of a neutron between bound state $i$ and $j$ is (see SI part III for derivation):

$$\Omega_{ij} = \frac{qm_n}{M\hbar} \frac{\omega_{ij}}{\omega} \vec{E} \cdot \int \bar{\psi}_i^*(\vec{r}) \vec{r} \bar{\psi}_j(\vec{r}') dV \qquad (7)$$

where $\omega$ and $\omega_{ij}$ are the microwave frequency and resonance frequency of the transition, $\vec{E}$ is the electric amplitude vector of the microwave. Initializing a neutron in the 1s state, the Rabi oscillation of the average neutron numbers in 1s and 1p states is shown in Fig. 3b. The Rabi oscillation is 2~3 orders of magnitude faster than its decay (here we only consider the decay from neutron absorption), allowing a pulse sequence of microwave control applied to the neutron qubit. Each pair of bound states following the selection rule of electric dipole transition has a transition matrix element, and the Rabi frequencies are generally on the order of magnitude of MHz with typical electric field intensity in experiments of kV/cm (for example, Ref. [37] applies an electric field of 3 kV/cm, and Ref. [38] applies a stronger electric field up to 500 kV/cm), as shown in Fig. 3c. The Rabi frequency has a negative relation with the nanocrystal radius, providing strong coupling between 1s and 1p states up to 5 MHz (Fig. 3d).

The above-studied neutron bound-state to bound-state transitions are mediated by microwave coupling to the mass of a charged quantum dot (the neutron does not couple directly to an electric field, but is coupled to the nuclide mass distribution of the quantum dot). It is also possible to excite a neutron bound-state to an unbound continuum scattering state by microwave, using the same principle. The ability to "launch" bound neutrons to a specific momentum state controlled by the microwave frequency, polarization, and the detailed morphology of the quantum dot may open new avenues for precision control of individual neutrons.

***Conclusion and Outlook***

In this work, we demonstrated with analytical models and numerical calculations that hydride nanoparticles can host neutron bound states with ~$\mu$eV binding energy, tens of nanometers extent, and ~ms lifetime. The weakly bound neutron state can be controlled by the electric field of a microwave with a Rabi frequency of ~MHz, to explore existing excited states. To trap neutrons into $\mu$eV bound states in experiments, the incident neutrons need to be cooled to milli-Kelvin temperature, which can be realized in the ultracold neutron (UCN) source [30]. The nuclear spins of hydrogens in the nanoparticle need to be polarized, which can be realized by the DNP technique. The NQD in a UCN bottle can be initialized to its ground state using a microwave with a frequency higher than the 1p state binding energy but lower than the 1s state binding energy. The microwave can deplete the neutron in excited states, and the ground states NQDs will accumulate. The population of NQDs in different molecular neutronic states can be read out by the same microwave pulse and counting the outgoing neutrons. That gives the population of neutrons in the excited states. The tens of nanometers spatial extent of the neutron bound states provides possibilities to have multi-qubit interaction. In comparison to the long-range electromagnetic interactions in Rydberg atom platforms generated by electrical dipole-dipole interactions [39], the effective interactions between two quantum particles in NQDs are generated by the wavefunction overlap. The system we propose with feasible state preparation and control opens up the possibility of exploring fundamental physics such as characterizing the strong nuclear interaction with high precision and exploring the quantum statistics of different particles, as well as developing certain quantum information processing applications [40].

## Acknowledgments

We thank Boning Li and Haowei Xu for their insightful discussion. This work was supported by the Office of Naval Research Multidisciplinary University Research Initiative Award No. ONR N00014-18-1-2497 and DTRA (Award No. HDTRA1-20-2-0002) Interaction of Ionizing Radiation with Matter (IIRM) University Research Alliance (URA).

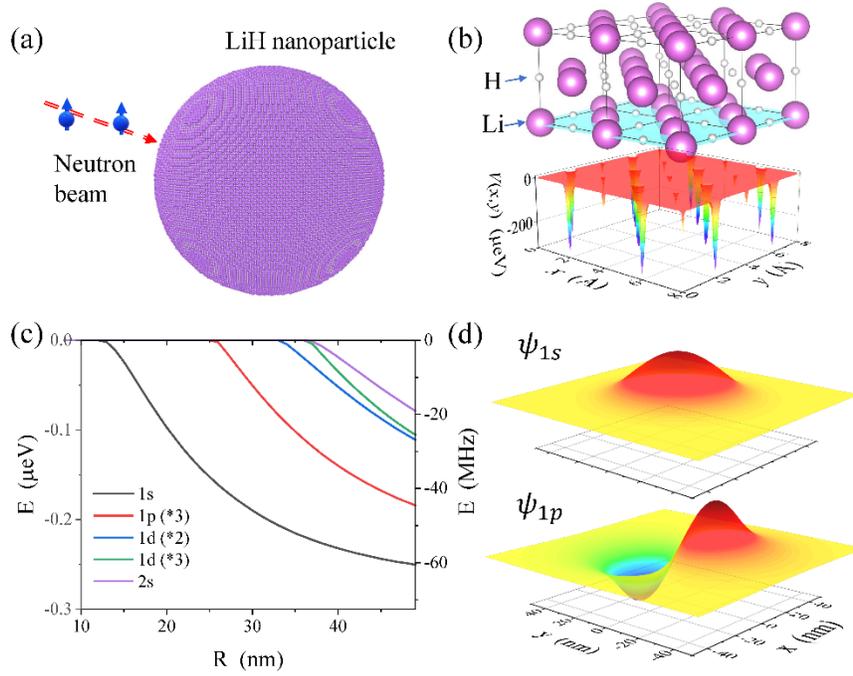

Figure 1: (a) Illustration of cold neutron bound states in 30 nm-radius LiH spherical nanocrystal. (b) Atomic structure (top) and nuclear force potential (bottom) of neutrons in LiH at zero temperature, where the hydrogen nuclear spins are fully polarized in an opposite direction with the neutrons. The nuclear force potential is smeared by the zero-point vibrations of nuclides to become a sum of picometer-lengthscale Gaussians and visualized on the (100) canyon plane. (c) Binding energies of molecular neutronic states as a function of nanocrystal radius. The energy levels are denoted as 1s, 1p (3-fold degeneracy), 1d (*2, 2-fold degeneracy), 1d (*3, 3-fold degeneracy), and 2s from low to high. (d) Real part of the average eigenfunction $\bar{\psi}(\vec{r})$ of 1s and 1p states when $R = 30$ nm plotted on a plane across the center of the sphere.

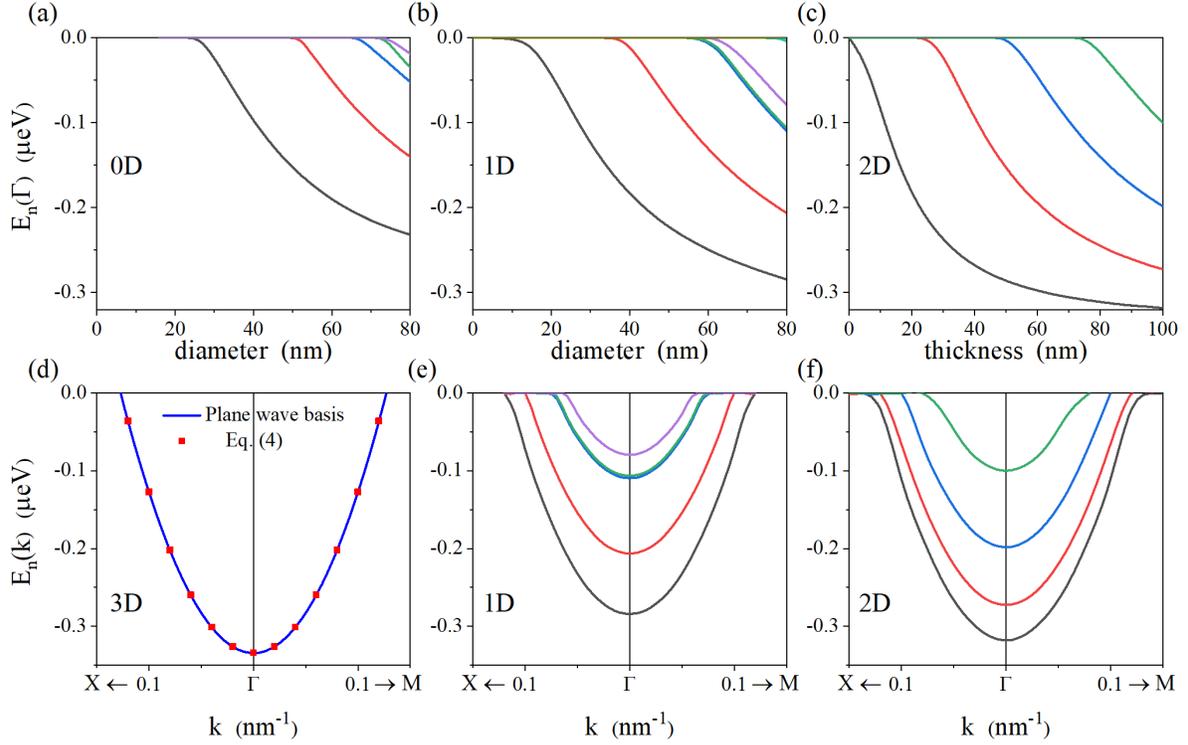

Figure 2: Neutron bound states in nanostructures with different dimensionality. Bound energy levels in LiH (a) zero-dimensional spherical nanoparticles, (b) one-dimensional cylindrical nanowire, and (c) two-dimensional thin film at Γ point as a function of the diameter of the nanoparticle, nanowire, and thickness of the thin film. (d) Neutron bound states band structure in perfect LiH crystal using Eq. (3) and a plane wave basis expansion method. (e) Neutron band structure in the 80 nm-diameter spherical nanowire and (f) 100 nm-thickness thin film. Throughout this figure, we assume hydrogen nuclear spins are fully polarized.

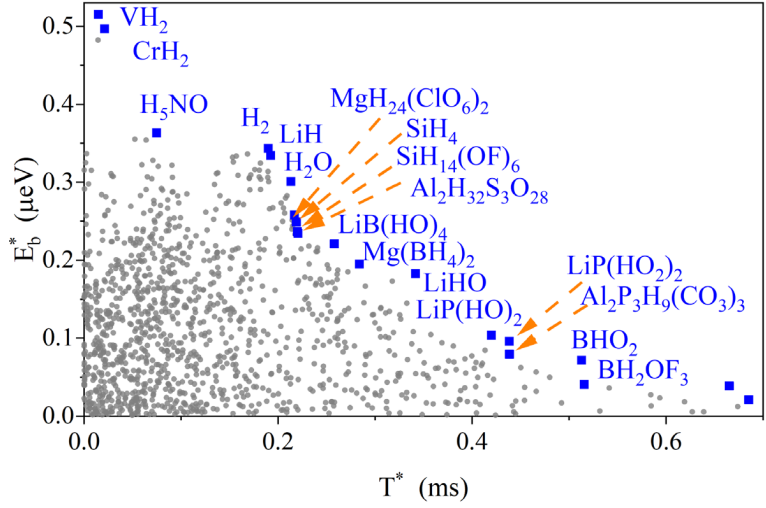

Figure 3: Binding energy $E_b^*$ and lifetime $T^*$ of molecular neutronic states in different perfect hydride crystals at zero temperature. The crystals are selected from 10,409 hydride systems from the materials project database, and crystals satisfying Pareto optimality with respect to exhibiting large binding energy and lifetime are denoted as blue squares.

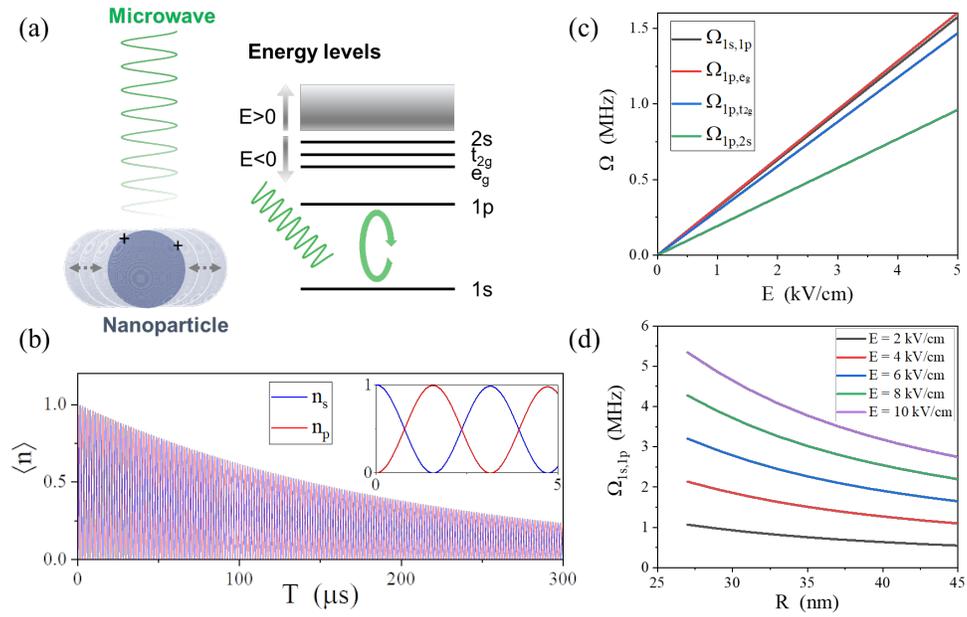

Figure 4: (a) Microwave control of neutron bound states in nanocrystals. The electric field of the microwave in resonance with 1s to 1p transition drives oscillations of charged nanocrystals and neutron states. (b) Rabi oscillation of neutronic 1s and 1p states ($R$ = 40 nm, $E$ = 1 kV/cm, $V_{\text{LiH}}$ = 1 V). The average population of neutrons in the 1s state ($n_s$) and 1p state ($n_p$) is plotted as a function of time. (c) Rabi frequency of different transitions as a function of microwave electric field ($R$ = 40 nm, $E$ = 1 kV/cm, $V_{\text{LiH}}$ = 1 V). (d) Rabi frequency of 1s to 1p transition as a function of nanocrystal radius and electric field ($V_{\text{LiH}}$ = 1 V).

# Supplementary Information: $\mu$eV-deep neutron bound states in nanocrystals

Hao Tang,[1] Guoqing Wang,[2,3] Paola Cappellaro,[2,3,4] and Ju Li[1,3,*]

[1]*Department of Materials Science and Engineering,*
*Massachusetts Institute of Technology, MA 02139, USA*
[2]*Research Laboratory of Electronics, Massachusetts Institute of Technology, Cambridge, MA 02139, USA*
[3]*Department of Nuclear Science and Engineering,*
*Massachusetts Institute of Technology, Cambridge, MA 02139, USA*
[4]*Department of Physics, Massachusetts Institute of Technology, Cambridge, MA 02139, USA*
(Dated: July 31, 2023)

## I. NEUTRON BOUND STATES

Here we describe the formalism of neutrons' low energy bound states in a nuclear force potential created by multiple nuclei. The Hamiltonian of the system is

$$H = -\frac{\hbar^2}{2m_n}\nabla^2 + \sum_i v_i(\vec{r}), \tag{1}$$

where $v_i(\vec{r})$ is the nuclear force potential energy of the $i$th nucleus to a neutron. Here, we employ the Green function formalism. The time-independent Schrodinger equation for an energy eigenfunction $\psi(\vec{r})$ can be written as:

$$\left[-\frac{\hbar^2}{2m_n}\nabla^2 + V(\vec{r})\right]\psi(\vec{r}) = E\psi(\vec{r}) = -\frac{\hbar^2\kappa^2}{2m_n}\psi(\vec{r}), \tag{2}$$

where we define $V(\vec{r}) \equiv \sum_i v_i(\vec{r})$ the wave vector $\kappa \equiv \frac{\sqrt{-2m_nE}}{\hbar}$ for bound states. The above equation is equivalent to:

$$\left(\kappa^2 - \nabla^2\right)\psi(\vec{r}) = -\frac{2m_nV(\vec{r})}{\hbar^2}\psi(\vec{r}). \tag{3}$$

The Green function of this equation satisfies:

$$\begin{cases} \left(\kappa^2 - \nabla^2\right)G(\vec{r},\vec{r}';E) = \delta(\vec{r}-\vec{r}') \\ \lim_{|\vec{r}-\vec{r}'|\to\infty} G(\vec{r},\vec{r}';E) = 0, \end{cases} \tag{4}$$

which gives

$$G(\vec{r},\vec{r}';E) = \frac{e^{-\kappa|\vec{r}-\vec{r}'|}}{4\pi|\vec{r}-\vec{r}'|}. \tag{5}$$

The energy eigenfunctions can then be formally expressed as:

$$\psi(\vec{r}) = -\int G(\vec{r},\vec{r}';E)\frac{2m_nV(\vec{r}')}{\hbar^2}\psi(\vec{r}')dr'^3, \tag{6}$$

which is Eq. (1) in the main text. The integral can be rewritten as a sum of local integrals around each nucleus:

$$\psi(\vec{r}) = -\sum_i \int_{\Omega_i} G(\vec{r},\vec{r}';E)\frac{2m_nv_i(\vec{r}')}{\hbar^2}\psi(\vec{r}')dr'^3, \tag{7}$$

where $\Omega_i$ is a spherical volume around the $i$th nucleus with a radius of the nuclear force range. We consider two cases: the first is $\vec{r}$ is not fm-scale close to any nuclei. As $G$ is a slowly varying function of $\vec{r}'$ that is approximately a constant at each nucleus' fm-scale neighborhood, it can be taken out of the integral, replacing $\vec{r}'$ by $\vec{R}_i$ (case 1 of main text Eq. (2)):

$$\psi(\vec{r}) = -\sum_i G(\vec{r},\vec{R}_i;E)\int_{\Omega_i}\frac{2m_nv_i(\vec{r}')}{\hbar^2}\psi(\vec{r}')dr'^3. \tag{8}$$

The second case is that $\vec{r}$ is ~fm close to the $j$th nucleus, so extracting $G$ out of the integral around $\Omega_j$ in Eq. (8) is invalid. Then we need to add a contribution from the $j$th nucleus' neighborhood (case 2 of main text Eq.(2)):

$$\psi(\vec{r}) = -\sum_{i\neq j} G(\vec{r},\vec{R}_i;E)\int_{\Omega_i}\frac{2m_nv_i(\vec{r}')}{\hbar^2}\psi(\vec{r}')dr'^3$$
$$-\int_{\Omega_j} G(\vec{r},\vec{r}';E)\frac{2m_nv_j(\vec{r}')}{\hbar^2}\psi(\vec{r}')dr'^3. \tag{9}$$

As we consider the low-energy case, this integral can then be simplified following the typical assumptions of the Fermi pseudopotential. We can introduce the scattering length, $b_i$, for low-energy neutron[1] whose real part is $\mathrm{Re}[b_i] \equiv \lim_{E\to0}\frac{1}{4\pi\bar\psi(\vec{R}_i)}\int_{\Omega_i}\frac{2m_nv_i(\vec{r}')}{\hbar^2}\psi(\vec{r}')dr'^3$, where $\bar\psi(\vec{r})$ is the average eigenfunction[1]:

$$\bar\psi(\vec{r}) = \frac{3}{4\pi D^3}\int_{|\vec{r}-\vec{r}'|<D}\psi(\vec{r}')dr'^3. \tag{10}$$

Here $D$ is a length that is far smaller than the interatomic distance but far larger than the force range of nuclear potential. This approximation can be taken because the interatomic distance is five orders of magnitude large than the force range of nuclear force. Eq. (8) and Eq. (9) can be simplified using the average eigenfunction and scattering length. The first case (Eq. (8)) gives:

$$\bar\psi(\vec{r}) \simeq \psi(\vec{r}) = -4\pi\sum_i G(\vec{r},\vec{R}_i;E)\mathrm{Re}[b_i]\bar\psi(\vec{R}_i)$$
$$= -\sum_i \frac{e^{-\kappa|\vec{r}-\vec{R}_i|}}{|\vec{r}-\vec{R}_i|}\mathrm{Re}[b_i]\bar\psi(\vec{R}_i), \quad \forall i(|\vec{r}-\vec{R}_i|\gg \mathrm{fm}), \tag{11}$$

where $\bar\psi(\vec{r})$ approximately equals $\psi(\vec{r})$ as the later is slowly varying at the length-scale $D$ when $\vec{r}$ is far from nuclei. The second case (Eq. (9)) gives:

$$\psi(\vec{r}) = -\sum_{i \neq j} \frac{e^{-\kappa|\vec{r}-\vec{R}_i|}}{|\vec{r}-\vec{R}_i|} \mathrm{Re}[b_i]\bar{\psi}(\vec{R}_i)$$
$$- \int_{\Omega_j} \frac{e^{-\kappa|\vec{r}-\vec{r}'|}}{|\vec{r}-\vec{r}'|} \frac{m_n v_j(\vec{r}')}{2\pi\hbar^2}\psi(\vec{r}')dr'^3, (|\vec{r}-\vec{R}_j| \sim \text{fm}). \quad (12)$$

The average eigenfunction around the nuclei can be derived by substituting this equation into Eq. (10). Using the slow varying condition again for the first term, we obtain

$$\bar{\psi}(\vec{r}) = -\sum_{i \neq j} \frac{e^{-\kappa|\vec{r}-\vec{R}_i|}}{|\vec{r}-\vec{R}_i|} \mathrm{Re}[b_i]\bar{\psi}(\vec{R}_i) - \frac{3}{4\pi D^3}$$
$$\times \int_{|\vec{r}_1-\vec{r}|<D} dr_1^3 \int_{\Omega_j} dr_2^3 \frac{e^{-\kappa|\vec{r}_1-\vec{r}_2|}}{|\vec{r}_1-\vec{r}_2|} \frac{m_n v_j(\vec{r}_2)\psi(\vec{r}_2)}{2\pi\hbar^2}. \quad (13)$$

In the second term, we can first do the integral $\int_{|\vec{r}_1-\vec{r}|<D} dr_2^3$, and as $\kappa D \ll 1$, the exponential term in the Green function approximates 1. As both $\vec{r}$ and $\vec{r}_2$ are fm-close to $\vec{R}_j$, their distance is also in fm scale, far smaller than $D$. The integral gives $2\pi D^2$, then the second term equals $-\frac{3}{2D}\mathrm{Re}[b_j]\bar{\psi}(\vec{R}_j)$. As $D \gg$ fm, this term is negligible compared to $\bar{\psi}$ itself, so only the first term remains. Summarizing the two cases gives the general expression of the average eigenfunction:

$$\bar{\psi}(\vec{r}) \simeq -\sum_{i, |\vec{r}-\vec{R}_i| \gg \text{fm}} \frac{e^{-\kappa|\vec{r}-\vec{R}_i|}}{|\vec{r}-\vec{R}_i|} \mathrm{Re}[b_i]\bar{\psi}\left(\vec{R}_i\right), \quad (14)$$

which is Eq. (3) in the main text. To get a closed set of equations, we set $\vec{r} = \vec{R}_i$ and denote $\bar{\psi}\left(\vec{R}_i\right)$ as $\bar{\psi}_i$:

$$\bar{\psi}_i + \sum_{j \neq i} \frac{e^{-\kappa|\vec{R}_i-\vec{R}_j|}}{|\vec{R}_i-\vec{R}_j|} \mathrm{Re}[b_j]\bar{\psi}_j = 0, \quad (15)$$

which is Eq. (4) in the main text.

In order to prove the existence of molecular neutronic states, we then need to show that the equations have non-zero solutions $\bar{\psi}_i$ with positive $\kappa$. In general cases, this can only be solved numerically. Here, we provide a simplified model that shows the existence of molecular neutronic states analytically.

Assuming there are nuclei on an infinite cubic lattice with a lattice constant of $a$, and each nucleus has a scattering length of $b$. We assume $\mathrm{Re}[b] \ll a$. Because of the periodicity of the lattice, the neutron eigenfunction follows the Bloch theorem. We consider the Bloch function at $\Gamma$ point, which means $\bar{\psi}_i$ on lattice points is a constant in the whole space. Then Eq. (15) gives:

$$1 + \frac{\mathrm{Re}[b]}{a} \sum_{(x_1,x_2,x_3) \neq (0,0,0)} \frac{e^{-\kappa a\sqrt{x_1^2+x_2^2+x_3^2}}}{\sqrt{x_1^2+x_2^2+x_3^2}} = 0. \quad (16)$$

We can easily see that the equation has a solution only when $\mathrm{Re}[b] < 0$. As $\frac{|\mathrm{Re}[b]|}{a} \ll 1$ and each term in the summation is less than 1, there must be a large number of lattice points contributing to the summation so that the second term can equal -1. That means the lattice is dense and the summation can be approximated as an integral:

$$1 + \frac{\mathrm{Re}[b]}{a} \int dx^3 \frac{e^{-\kappa a|x|}}{|x|} = 0. \quad (17)$$

The integral can be done analytically, giving $\frac{4\pi}{\kappa^2 a^2}$. The solution of the binding energy is then

$$E_b^* = \frac{\hbar^2 \kappa^2}{2m_n} = -\frac{2\pi\hbar^2 \mathrm{Re}[b]}{m_n a^3}. \quad (18)$$

Similarly, if a unit cell contains multiple nuclei in an infinite crystal, the condition that allows this equation to have a solution is $\sum_\alpha n_\alpha \mathrm{Re}[b_\alpha] < 0$, where $n_\alpha$ and $b_\alpha$ are the number and scattering length of the $\alpha$th type of nucleus in a unit cell. The solution of the binding energy is then

$$E_b^* = \frac{\hbar^2 \kappa^2}{2m_n} = -\frac{2\pi\hbar^2 \sum_\alpha n_\alpha \mathrm{Re}[b_\alpha]}{m_n V_{u.c.}}, \quad (19)$$

which is Eq. (5) in the main text ($V_{u.c.}$ is the unit cell volume). The order of magnitude of $E_b$, considering hydrogen-storage materials, is 0.1~ 1 $\mu$eV. The dispersion relation can also be obtained as

$$E(k) = \frac{\hbar^2}{2m_n} \left( \frac{4\pi \sum_\alpha n_\alpha \mathrm{Re}[b_\alpha]}{V_{u.c.}} + k^2 \right). \quad (20)$$

Using this equation, we can evaluate the mass gain of the LiH crystal when trapping neutrons. The Brillouin zone volume with $E(k) < 0$ is $V_{B.Z.} = \frac{4\pi}{3} \left( -\frac{4\pi \sum_\alpha n_\alpha \mathrm{Re}[b_\alpha]}{V_{u.c.}} \right)^{3/2}$. Assuming the neutron-neutron interaction in the molecular neutronic states is negligible, the neutron mass density in the material is $\rho_n = \frac{m_n V_{B.Z.}}{(2\pi)^3}$. Then, the percentage mass gain discussed in the main text is:

$$\frac{\rho_n}{\rho_{LiH}} = \frac{4m_n}{3\sqrt{\pi}\rho_{LiH}} \left( -\frac{\sum_\alpha n_\alpha \mathrm{Re}[b_\alpha]}{V_{u.c.}} \right)^{3/2} \times 100\%. \quad (21)$$

Using the data of LiH crystal ($\rho_{LiH} = 780kg/m^3$, $n_{Li} = n_H = 4$, $\mathrm{Re}[b_{Li}] = -2.22$ fm ($^7$Li, unpolarized), $\mathrm{Re}[b_H] = -18.33$ fm ($^1$H, fully polarized), $V_{u.c.} = 68.09$ Å$^3$, $m_n = 1.675 \times 10^{-27}$ kg), we get a mass gain of $6.78 \times 10^{-6}\%$.

As an infinite crystal can host molecular neutronic states, we can conclude that a finite NQD can host such states as long as its size is sufficiently large. Then, we aim to numerically solve the molecular neutronic states for finite-size NQD. In order to obtain a manageable numerical problem, we do a coarse-graining again by turning Eq. (15) into an integral equation using again the continuum approximation:

$$\bar{\psi}(\vec{r}) + \int_{\Omega_{n.c.}} dr'^3 \frac{e^{-\kappa|\vec{r}-\vec{r}'|}}{|\vec{r}-\vec{r}'|} \frac{\sum_\alpha n_\alpha \mathrm{Re}[b_\alpha]}{V_{u.c.}} \bar{\psi}(\vec{r}') = 0, \quad (22)$$

where $\Omega_{n.c.}$ is the nanocrystal region. The equation is then discretized by a coarse grid lattice $\mathbf{r}_i = (x_i, y_i, z_i)a_0$ with integer numbers of $x_i, y_i, z_i$. Then, the equation turns into:

$$\bar{\psi}(\mathbf{r}_i) + \frac{a_0^3 \sum_\alpha n_\alpha \mathrm{Re}[b_\alpha]}{V_{u.c.}} \sum_{j \neq i} \frac{e^{-\kappa r_{ij}}}{r_{ij}} \bar{\psi}(\mathbf{r}_j) = 0, \quad (23)$$

where $r_{ij} = |\mathbf{r}_i - \mathbf{r}_j|$, and grid separation $a_0$ is set as $R/10$, one tenth of the nanocrystal radius. All grid points within the nanocrystal sphere are included in Eq. (23) and are solved as an eigenvalue problem with eigenvalue $\kappa$ and eigenvector $(\bar{\psi}(\mathbf{r}_1), \bar{\psi}(\mathbf{r}_2), \cdots)$. Our numerical test shows that this $a_0$ gives good numerical convergence of binding energy to three effective digits accuracy.

Finally, we reexamine the applicability of Fermi pseudopotential in molecular neutronic states. Here, we prove that $\lim_{E \to 0} \frac{1}{4\pi \bar{\psi}(R_i)} \int_{\Omega_i} \frac{2m_n v_i(\vec{r})}{\hbar^2} \psi(\vec{r}) dr'^3$ converges to the same number regardless of whether $E$ approaches zero from a positive or negative direction. That validates our usage of the scattering length from neutron scattering data to molecular neutronic states. As the force range of $v_i(\vec{r})$ is far smaller than $1/\kappa$, the eigenfunction near a nucleus is accurately approximated as an $s$-wave. We denote the wave function around $R_i$ as $\psi(r)$ and $u(r) \equiv r\psi(r)$, where $r$ is the distance from the nucleus. The eigenvalue equation is:

$$-\frac{\hbar^2}{2m_n} \frac{d^2 u}{dr^2} + v_i(r) u(r) = E u(r). \quad (24)$$

At $r$ larger than the force range $R_0$, the potential $v_i(r) = 0$, then we have $u''(r) = 0$ given $E \to 0$. That means the $u(r)|_{r > R_0} = k(r + c_0)$ is a linear function. The average eigenfunction $\bar{\psi}_i = \frac{3}{4\pi D^3} \int_0^D u(r) 4\pi r dr \simeq k$ (as $c_0, R_0 \ll D$). We can then integrate Eq. (24) multiplied by $r$ from $r = 0$ to $r = R_0$:

$$-\frac{\hbar^2}{2m_n} \int_0^{R_0} dr \frac{d^2 u}{dr^2} r + \int_0^{R_0} v_i(r) u(r) r dr = 0. \quad (25)$$

That gives:

$$-\frac{\hbar^2}{2m_n} \left[ r \frac{du}{dr} - u(r) \right]_0^{R_0} + \frac{1}{4\pi} \int_{\Omega_i} v_i(\vec{r}) \psi(\vec{r}) dr = 0. \quad (26)$$

As $u(0) = 0$, $R_0 \frac{du}{dr}|_{R_0} - u(R_0) = -kc_0$, we obtain

$$\frac{\hbar^2}{2m_n} c_0 \bar{\psi}_i + \frac{1}{4\pi} \int_{\Omega_i} v_i(\vec{r}) \psi(\vec{r}) dr'^3 = 0. \quad (27)$$

That proves the expression $\frac{1}{4\pi \bar{\psi}(R_i)} \int_{\Omega_i} \frac{2m_n v_i(\vec{r})}{\hbar^2} \psi(\vec{r}) dr'^3$ approaches a constant of $-c_0$, which is defined as $\mathrm{Re}[b_i]$, when $E$ approaches zero. Here, neglecting the $E$-term applies as long as $\sqrt{2m_n|E|}/\hbar \ll \mathrm{fm}^{-1}$ regardless of the sign of $E$, so the low-energy scattering state and bound state have the same scattering length.

## II. NEUTRON ABSORPTION LIFETIME AND MATERIALS SCREENING

In addition to the binding energy, it is also important to evaluate other effects that might limit the lifetime of molecular neutronic states. In the Hamiltonian in Eq. (1), we did not consider electromagnetic radiation, which leads to the radiative neutron absorption process where a neutron drops into a MeV deep bound state and emits a $\gamma$-ray photon. This neutron absorption process makes the neutron bound state unstable. The low-energy neutron absorption rate is proportional to the Fermi contact, with a constant of low-energy absorption cross-section $\sigma_a(E_k)$ multiplied by neutron velocity $\sqrt{2E_k/m_n}^2$. Summing over all nucleus, the absorption rate of a molecular neutronic state in a uniform perfect crystal is:

$$\frac{d\Gamma}{dt} = \frac{1}{T^*} = \sum_i |\bar{\psi}_i|^2 \lim_{E_k \to 0} \sigma_a^i(E_k) \sqrt{\frac{2E_k}{m_n}}$$
$$= \lim_{E_k \to 0} \frac{1}{V_{u.c.}} \sum_\alpha n_\alpha \sigma_a^\alpha(E_k) \sqrt{\frac{2E_k}{m_n}}, \quad (28)$$

which is Eq. (5) in the main text. In the second line, we use the condition that $\bar{\psi}$ is normalized and constant in space.

If the nanocrystal has a finite size, the argument based on the uniform wave function is no longer valid. Interestingly, the product of binding energy and absorption rate are upper bounded. The binding energy can be rewritten as the opposite of the summation of potential energy and kinetic energy:

$$E_b = -\frac{2\pi\hbar^2}{m_n} \sum_i \mathrm{Re}[b_i] |\bar{\psi}_i|^2 - \langle \dot{E}_k \rangle \leq -\frac{2\pi\hbar^2}{m_n} \sum_i \mathrm{Re}[b_i] |\bar{\psi}_i|^2,$$
$$(29)$$

where $\langle \dot{E}_k \rangle$ is the average kinetic energy, sum over $i$ goes through all unit cells $m$ and all the atom types $\alpha$ in each unit cell. As $\bar{\psi}$ is slowly varying in the scale of a unit cell, the inequality approximates:

$$E_b \leq -\frac{2\pi\hbar^2}{m_n} \sum_m |\bar{\psi}_m|^2 \sum_\alpha n_\alpha \mathrm{Re}[b_\alpha], \quad (30)$$

where $\bar{\psi}_m$ is the average eigenfunction in the $m$th unit cell. The absorption rate, by this notation, can be written as:

$$\frac{1}{T} = \sum_m |\bar{\psi}_m|^2 \lim_{E_k \to 0} \sum_\alpha n_\alpha \sigma_a^\alpha(E_k) \sqrt{\frac{2E_k}{m_n}}. \quad (31)$$

We therefore have

$$E_b T \leq -\frac{2\pi\hbar^2}{m_n} \frac{\sum_\alpha n_\alpha \mathrm{Re}[b_\alpha]}{\lim_{\alpha \to 0} \sum_\alpha n_\alpha \sigma_a^\alpha(E_k) \sqrt{\frac{2E_k}{m_n}}} = E_b^* T^*,$$
$$(32)$$

which is Eq. (6) in the main text. It is worth noticing that the inequality can be transformed into an inspiring form, which is similar to the time-energy uncertainty principle. The low energy absorption cross section is related to the imaginary part of the scattering length[2]:

$$\sigma_a^\alpha(E_k) = \frac{4\pi \, \mathrm{Im}[b_\alpha]}{k} = 4\pi \, \mathrm{Im}[b_\alpha]\sqrt{\frac{\hbar^2}{2m_n E_k}}. \quad (33)$$

Substituting this back into Eq. (32), we derive

$$E_b T \leq -\frac{\hbar}{2}\frac{\sum_\alpha n_\alpha \, \mathrm{Re}[b_\alpha]}{\sum_\alpha n_\alpha \, \mathrm{Im}[b_\alpha]}. \quad (34)$$

We see that compared to the time-energy uncertainty principle, the upper bound has an amplification of $\frac{\mathrm{Re}[b_\alpha]}{\mathrm{Im}[b_\alpha]}$, which is $10^6$ for polarized hydrogen atoms.

In the materials screening, we select crystal structures from all available structures in the Materials project containing hydrogen. We first exclude structures not stable and structures containing elements heavier than La (because they generally have excessively high absorption cross sections and cannot give a reasonable lifetime), and then calculate $E_b^*$ and $T^*$ by Eq. (5) in the main text. The values of $b_i$ are extracted from ref.[2], where we assume that H nuclei are fully polarized, and all other nuclei are non-polarized. For most elements, we assume a natural abundance of isotopes, except for a few elements: we assume Li, B, Cl, and Se are purified as $^7$Li, $^{11}$B, $^{37}$Cl, $^{80}$Se, because these isotopes are naturally abundant and the purification significantly improves lifetime in some compound.

## III. MICROWAVE CONTROL

The microwave applies an electric field on a charged nanocrystal with a mass of $M = \frac{4\pi}{3}R^3\rho$ and an electric charge of $q$. If the electric field is

$$\vec{E}(t) = \vec{E}_0 \sin \omega t. \quad (35)$$

Then the coordinates $\vec{R}$ of the center of the charged nanocrystal is

$$\vec{R}(t) = -\frac{q\vec{E}_0}{M\omega^2}\sin \omega t. \quad (36)$$

The time-dependent Schrodinger equation $\hat{H}_{\vec{R}(r)}\psi = i\hbar\frac{\partial \psi}{\partial t}$ gives:

$$c_m'(t) + \sum_n c_n(t)\dot{\vec{R}}(r)\cdot\langle\psi_{m,R(t)}|\nabla_R|\psi_{n,R(t)}\rangle = \frac{1}{i\hbar}c_m(t)E_m, \quad (37)$$

where $c_m(t)$ is the wave function in the energy representation and Schrodinger picture: $|\psi(t)\rangle = \sum_m c_m(t)|\psi_{m,R(t)}\rangle$, where $\hat{H}_{R(t)}|\psi_{m,R(t)}\rangle = E_m|\psi_{m,R(t)}\rangle$. Transforming to the interaction picture, we derive

$$\frac{d}{dt}c_m^I(t) + \sum_{n\neq m} c_n^I(t)\Omega_{mn}e^{\frac{i}{\hbar}(E_m - E_n)t}\cos \omega t = 0, \quad (38)$$

where the Rabi frequency equals:

$$\Omega_{mn} = \frac{qm_n\vec{E}_0}{M\hbar}\frac{E_m - E_n}{\hbar\omega}\cdot\langle\psi_{m,R(t)}|\vec{r}|\psi_{n,R(t)}\rangle. \quad (39)$$

As the matrix element does not depend on $t$, so we write it as $\langle\psi_m|\vec{r}|\psi_n\rangle$ in Eq. (7) in the main text. In the derivation, we used the fact that $\psi_{n,R(t)}$ is a function of $\vec{r}-\vec{R}(t)$. so $\nabla_R\psi_{n,R(t)} = -\nabla_r\psi_{n,R(t)}$. Then,

$$\begin{aligned}
\langle\psi_{m,R}|\nabla_R|\psi_{n,R}\rangle &= -\langle\psi_{m,R}|\nabla_r|\psi_{n,R}\rangle \\
&= -\frac{m_n}{\hbar^2}\langle\psi_{m,R}|[\vec{r},\hat{H}]|\psi_{n,R}\rangle \\
&= (E_m - E_n)\frac{m_n}{\hbar^2}\langle\psi_{m,R}|\vec{r}|\psi_{n,R}\rangle.
\end{aligned} \quad (40)$$

This completes the derivation from Eq. (37) to the Rabi frequency expression Eq. (39).

Numerically, we calculate the matrix element by discretizing the integral on the same lattice as described in section 1. The integral is done on a cubic region whose center is the sphere center of the nanocrystal and the length of a side is 3 times of the sphere radius. The numerical test shows that the box size of the integral gives good numerical convergence of the matrix element to three effective digits accuracy.